\documentstyle[preprint,prl,aps,12pt]{revtex}

   \def \ybco
{YBa$_{2}$Cu$_{3}$O$_{7-x}$}

\begin{document}

\title {\bf Quasiparticle contribution to heat carriers relaxation time \\ in
DyBa$_2$Cu$_3$O$_{7-x}$ from heat diffusivity measurements.}

\author{S. Dorbolo(+) and M. Ausloos(\dag)\\ S.U.P.R.A.S.\\ (+) Institut
d'\'{e}lectricit\'{e} Montefiore B28\\ (\dag) Institute of Physics B5\\
University of Li\`{e}ge, B-4000 Li\`{e}ge\\ Belgium}

\maketitle

\begin{abstract} It is shown that the controversy on phonons or electrons being
the  most influenced heat carriers below the critical temperature of  high-T$_c$ superconductors can be
resolved. Electrical and thermal properties of the same DyBa$_2$Cu$_3$O$_{7-x}$
monodomain have been measured for two highly different oxygenation levels.  While the
oxygenated sample DyBa$_2$Cu$_3$O$_{7}$ has very good superconducting
properties
($T_c=90$ K), the DyBa$_2$Cu$_3$O$_{6.3}$ sample exhibits an insulator behavior. A careful
comparison between measurements of the {\bf thermal diffusivity} of both samples allows us to extract the electronic
contribution. This
contribution to the relaxation time of heat carriers is shown to be large below
$T_c$ and more sensitive to the superconducting state than the phonon
contribution.

\vfill P.A.C.S. : 60 +w, 17.15 jf, 44.90 +c \end{abstract}

\setcounter{section}{0}

\section{Introduction} As far as the temperature dependence of
the relaxation time of
{\it heat} carriers in high-$T_c$ superconductors (HTS) is concerned,
different regions
can be highlighted. At high temperature, the main process is due to phonon
scattering\cite{ziman,zeini}.  At very low temperature, the different
scattering
processes are decoupled as shown by fast heat response measurements
\cite{maneval}. At intermediate temperatures, below the critical temperature
$T_c$, a bump characterizes the thermal conductivity $\kappa$ data. This bump is linked to an increase of the mean free
path $\ell$ of the heat carriers or in other words to an increase of their
relaxation time $\tau$. A major controversy subsists on the origin of such a
behavior, thus about the thermal transport mechanisms themselves below the critical
temperature $T_c$ \cite{zeini,SUST,zhang}. Two main models collide. The aim
of this work is
to determine the nature of the carrier responsible for the increase of the
relaxation time $\tau$ of heat carriers below $T_c$ and solve the puzzle.

\par On one hand, W\"{o}lkhausen et al.\cite{wolk} have proposed a so-called
phonon model in order to explain the bump in $\kappa$ below the critical
temperature $T_c$.  This model which introduces 7 parameters
describes different
relaxation time processes. Phonons are thought to be less scattered
by electrons
since these condense into Cooper pairs.  If the main scattering process below
$T_c$ is taken to be the phonon-electron process, the calculation of $\kappa$
does reproduce
the bump observed.  One can nevertheless wonder whether the origin of
the bump in
such a scheme is due to the reduced number of scatterers or to their mobility
increase.

\par On the other hand, the electronic model\cite{yu,houssa}
in which the
main heat carriers are charges claims that the bump reflects the {\it mean free
path} increase of electrons, below $T_c$,  because the number of
electron-electron scattering processes decreases. Indeed below $T_c$, normal
electrons (quasiparticles) should be less often scattered by others because of
the existence of Cooper pairs (which cannot carry entropy in the condensed
state). This argument is especially interesting in HTS in which the gap symmetry is of $d$-wave
type\cite{kirthley}, thus characterized by lines of nodes.  Along these
the gap is
zero\cite{scalapino}, whence the normal electrons  {\it density}
always remaining finite
even at very low temperature.

\par From the $electrical$ transport point of view, a quasiparticle relaxation
time $\tau_{e}$ can be deduced from microwave loss measurements, as
done by Bonn
et al. \cite{bonn} on \ybco \ monocrystal. Another way to determine
$\tau_{e}$  through 
the magneto-transport experiments was made by  Krishana
et al. \cite{krishana} on a Zn-doped YBa$_2$Cu$_3$O$_{7}$.  They measured the thermal Hall effect, also called
Righi-Leduc effect, i.e. the temperature gradient
orthogonal to the plane defined by the heat flux and an external
magnetic field.
In so doing, the {\it electronic contribution} is certainly selected since electrons
are deviated by the magnetic field $B$. Interestingly, these Righi-Leduc effect measurements scale with the
measurements of
microwave losses 
\cite{bonn2} giving a correspondence between electrical and thermal
measurements
of the relaxation time in presence of a magnetic field.  

\par However such a magneto-thermal process implies high magnetic fields which
disturb the quasi-particle ($qp$) system and blur interpretation at zero field.  In fact, not only the
distinction
between $qp$ and vortex contributions is far from obvious,\cite{MASS} but also
the existence or not of Landau levels has been argued upon and is still an open
question \cite{zhangand6}.  Moreover in \cite{krishana}, the relaxation time is
found by achieving extrapolation to zero-field; that is known to be very
delicate in view of the complicated ($B$, $T$) phase diagram of
HTS\cite{blatter}.

\section{Method}

\par In contrast we propose a {\it direct comparative method} of electrical and
thermal measurements in order to deduce whether phonons or electrons are
responsible for the increase of the relaxation time of heat carriers
below $T_c$.
The measurement of the thermal diffusivity coefficient $\alpha$ allows one a
$direct$ access to the relaxation time of heat carriers indeed. Let it be
recalled that  $\alpha$ reads \cite{cahill} \begin{equation}
\alpha=\frac{1}{3}v \ell =\frac{1}{3}v^2 \tau \end{equation} 
if  $2  \pi \nu \tau \gg 1$, where $\nu$ is the
phonon or electron change in frequency,
and $v$ is the
average speed of the carriers.  For electrons, the change of energy close to $T_c$ is of the order of the gap energy; the change in frequency is $\nu \simeq 10^{13}$ Hz (for Dy-123) while $\tau \simeq 10^{-12}$ s \cite{bonn}.  Claiming
without lack of much
generality that $v$
is constant in a $50$ K range around $T_c$, the thermal diffusivity can be
assimilated to the relaxation time. Moreover, $\alpha$ can be decomposed as
follows \begin{equation} \alpha=\frac{\kappa}{d c}=\frac{\kappa_e
+\kappa_{ph}}{d
c}=\frac{\kappa_e}{d c} +\frac{\kappa_{ph}}{d c}=\alpha_e + \alpha_{ph} \propto
\tau_e +\tau_{ph} \end{equation} where $\kappa$ is the thermal
conductivity, $c$
the specific heat ($c_e \ll c_{ph}$) and $d$ the density.  The
subscripts $e$ and
$ph$ represent the electronic and the phononic contribution in each
quantity.

\par A home made apparatus has been constructed to measure $\alpha$
{\it directly}, $together$ and $simultaneously$ with the thermoelectric
power $S$ and
$\kappa$ \cite{epj,kappaexp}.  The same thermocouples are used to measure those three
physical parameters.  The thermal diffusivity was found to be described
in a log-log plot by two straight lines crossing at $T_c$
\cite{batonrouge} with
the power law exponent larger below than above $T_c$.  Fig.1 shows a sketch of
the behavior of $\alpha$ versus temperature on a log-log plot for further reference. The normal
state is designated by the {\bf +} sign  and the superconducting state by the
{\bf -} sign.  The excess contribution due to the superconducting phase is found by extrapolating the normal state value below the critical
temperature (grey area in Fig.1). The thermal diffusivity below $T_c$ can be written as
\begin{equation}\label{alphasup}
\alpha^{-}=\alpha^{-}_{e}+\alpha^{-}_{ph}=\alpha^{s}_{e}
+\tilde{\alpha}^{+}_{e}+\alpha^{s}_{ph}+\tilde{\alpha}^{+}_{ph} \end{equation} where $\alpha^{s}_{e}$ et
$\alpha^{s}_{ph}$ are the true superconducting (s) contributions of electrons
and phonons below $T_c$ respectively and the tilde quantities are the normal state ones extrapolated below $T_c$.  Our method consists in evaluating
the extrapolated value of $\alpha^{+}_{ph}$ below $T_c$, i.e. $\tilde{\alpha}^{+}_{ph}$, in
order to remove this term from Eq.(\ref{alphasup}), and to have only
$\alpha_{e}^{-}+\alpha_{ph}^{s}$ to analyze. This quantity is then $compared$ to
the quasiparticle relaxation time obtained by microwave losses measurements from
\cite{bonn} in order to $estimate$ $\tau_e$, in our notations $\tau_e^{-}$, thus $\alpha_e^{-}$. In
such a method,
no model is {\it a priori} assumed concerning the nature of heat transport
carriers responsible for the increase of $\tau$ (in other words,
$\alpha$) below
the critical temperature.

\par The various contributions to $\alpha$ are obtained by
comparing the
thermal properties of a Dy-123 monodomain before and after
deoxygenation as in ref.\cite{cohn} for polycrystaline superconductors. In so
doing, the thermal diffusivities of the sample should strictly
behave as those
of a superconductor in the well oxygenated case ($\alpha_{ox}$) or as those of an insulator
in the
deoxygenized case ($\alpha_{deox}$) because the density of holes is then largely
decreased.  A main argument is that the geometry is $conserved$ and the phonon
spectrum quasi undisturbed, such that any spurious effect can be easily taken
into account
by rescaling at high temperature.  Is so doing, $\tilde{\alpha}_{ph}^{+}$ can then be assimilated to the thermal diffusivity of the deoxygenated sample $\alpha_{deox}$.

\section{Synthesis and experiments}
The simple domain synthesis is described in
ref.\cite{dysynthese}.  A bar was cut out from the single grain. Its dimension
was 15 $\times$ 2 $\times$ 2 mm$^3$.  The electrical resistance $R(T)$,
thermal conductivity,
thermoelectric power $S(T)$ (not shown here), thermal diffusivity have been
measured and indicate the features expected from a good sample and allow for
further work.
The sample was then deoxygenized. This operation was controlled by TGA
(Thermo-gravimetric analysis). The weight of the sample was
continuously recorded
while it was heated in a furnace up to 850 ¡C. The composition of the sample was
thereby determined since the loss of weight was essentially due to the loss of
oxygen. Starting from a fully oxygenated sample DyBa$_2$Cu$_3$O$_7$
as seen from
$R(T)$, $S(T)$ and then $T_c$, a DyBa$_2$Cu$_3$O$_{6.3}$ stoichiometry was found
after TGA in
the present case.

\par The resistivity measurements have been achieved in a PPMS (Physical
Properties Measurement System from Quantum Design).  The thermal
conductivity and
thermoelectric power on one hand, and the thermal
diffusivity on the 
other hand, have been simultaneously measured following the steady state
method and the pulse method 
ref.\cite{epj} respectively, using the home made set up described in ref.\cite{kappaexp}.  

\section{Results} \par To begin with, the normalized resistance of
the oxygenated $R_{oxy}$
and deoxygenized samples $R_{deox}$ are compared on Figs.2 (a-b) respectively.  The
electrical behaviors are quite opposite to each other. On one hand, the
oxygenated sample, Fig.2(a), is a very good superconductor with a very sharp
transition. On
the other hand, the deoxygenized sample, Fig.2(b), exhibits a semiconducting (nearly insulating) behavior.
For the latter sample, its N\'{e}el temperature ($T_N=230$ K) has been revealed
by applying
a small magnetic field.  Notice that the maximum at $T\simeq 45$ K is slightly field dependent.  The origin of such a maximum is unknown.  It might be due to the existence of a spin glass or canted spin phase extending in the AF insulating phase.  The curve shape reminds of that found in GMR materials \cite{gmr}.  This maximum limits the subsequent analysis to the above $50$ K range.  It seems therefore quite reasonable to claim that two
different regions of the ($T_c$,$x$) phase diagram \cite{varma} are
investigated,
and that the electronic contribution to the electrical transport is wholly
suppressed in the deoxygenated sample.

\par The thermal conductivity is shown in Fig. 3 for both cases. Since those
results have been obtained for the same sample, the geometric factor does not
influence the absolute values of the measurements.  For the oxygenated sample ($\bullet$), the bump below the
critical temperature in the thermal conductivity $\kappa_{oxy}$ is well defined.  A
sharp minimum occurs
near $T_c$ at  $90$ K. This feature is not found when the sample is
deoxygenized ($\circ$)
: the bump disappears and the thermal conductivity $\kappa_{deox}$ slope remains quite
unchanged, smoothly increasing with $T$ up to high temperatures.  In the normal state, the thermal conductivity increases in both case, thus behaving like for a glass.  Notice the flattening of $\kappa_{deox}$ curve below $50$ K which might as for the resistivity due to some magnetic effects.

\par The total phonon contribution to the thermal conductivity $\kappa_{ph,t}$ can be written as \cite{ziman2}
\begin{equation}\label{kpa}
\kappa_{ph,t}^{-1}=\kappa_{ph}^{-1}+\kappa_{ph,b}^{-1}+\kappa_{ph,i}^{-1}
\end{equation} where $\kappa_{ph}$, $\kappa_{ph,b}$ and $\kappa_{ph,i}$ are the contributions due to the scattering of phonons by phonons, of phonons by boundaries and of phonons by impurities respectively.  The slope of the heating conductivity of the deoxygenated sample is higher than in the oxygenated case.  This can be explained by the thermal treatment used in order to decrease the oxygenation level by the TGA process.  The terms $\kappa_{ph,b}$ and $\kappa_{ph,i}$ are then slightly modified in Eq.(\ref{kpa}).  As far as the term $\kappa_{ph}$ is concerned, calculations for both oxygenated and deoxygenated samples show that their phonon spectra are very close to each other \cite{tyuterev}.  The authors of Ref.\cite{tyuterev} also claim that just a few modes can be distinguish frequency spectra of the oxygenated and deoxygenated case.  Consequently they are suitable as the reference phonons for the purpose of superstructure characterization, but the thermal transport by phonons should be only slightly affected.  Therefore, $\tilde{\alpha}_{ph}^{+}$ can be interpreted as the thermal diffusivity of the deoxygenized sample  $\alpha_{deox}$ : it is a purely phononic
contribution.  Nevertheless we stress that the analysis is valid only above $50$ K up to $T_c$.

\par The measurements of both (normalized) thermal diffusivities are plotted in
Fig.4 in a log-log plot.  The behavior of the thermal diffusivity of the
oxygenated sample $\alpha_{oxy}$ is typical of high-T$_c$ superconductors
\cite{kato,calzona,castel,wu,ikebe}. Straight lines are found in the
normal state
and in the superconductor state in the temperature range of interest. The
characteristic power law exponents are found to be $-0.77$ and $-1.5$ in the
normal and in the superconductor state. Those exponents can be explained
along the lines of the electronic theory previously reported in
\cite{batonrouge}.
This behavior is not found in the thermal diffusivity $\alpha_{deox}$ of
the deoxygenized sample.  Note that both $\alpha$'s are superposed to each other in
the normal state exactly like in the sketch of Fig.1.  In order to $numerically$ substract $\alpha_{deox}$ from $\alpha_{oxy}$, $\alpha_{deox}$ has been smoothened by filtering high frequencies in the Fast Fourier Transform spectrum of the signal.  In Fig.5,
$\alpha_{oxy}-\alpha_{deox}$ ($\bullet$) is shown versus the reduced
temperature $\epsilon = |T-T_c|/T_c$.  It is 
obvious from the graph that we can write 
\begin{equation}\label{quant}
\alpha_{oxy}-\alpha_{deox}\equiv \alpha^{-}-\tilde{\alpha}_{ph}^{+}=\alpha_{e}^{-}+\alpha_{ph}^{s} 
\end{equation} For comparison, the $qp$
relaxation time from Bonn et al. \cite{bonn} is also shown as circles
($\circ$), which are rescaled by a multiplicative factor.  The agreement is remarkable.  The quantity of Eq.(\ref{quant}) is proportionnal to $\tau_{e}$, the $qp$ relaxation time found by
microwave loss measurements.  This shows that the $electronic$ term $\alpha_{e}^{-}$ is unambiguously the dominant term in Eq.(\ref{alphasup}).

\section{Conclusions} In conclusion, a comparison between
electrical and thermal measurements of an oxygenated and desoxygenated sample allows one to determine the nature
of thermal
carriers in high T$_c$ superconductor ceramics. Direct measurements of the thermal diffusivity $\alpha$ on a DyBa$_2$Cu$_3$O$_{7-x}$ sample 
are reported for $x=0$ and $x=0.7$.  Such a direct measurement method of $\alpha$ leads to an unambiguous
result without the use of any magnetic field and without any {\it a priori}
assumption about the theoretical model used for describing the scatterers :
phonons or electrons. Electrons are found to be \underline{the most influenced} heat
carriers below the critical temperature. Phonons are only background carriers.

\indent \indent \pagebreak \par {\large \bf Acknowledgements} \\
\indent Part of
this work has been financially supported by the ARC 94-99/174 contract of the
Ministry of Higher Education and Scientific Research through the University of
Li\`{e}ge Research Council.  S. Dorbolo has benefited from a FRIA research
fellowship. We would like also to thank Prof. H.W. Vanderschueren for
the use of
MIEL equipments.  We want also to thank Prof. J.P. Maneval (ENS, Paris) and Pr.
E. Silva (U. Roma 3, Roma) and referees for their comments.

e-mail: \\ (+) S.Dorbolo@ulg.ac.be \\ (\dag) Marcel.Ausloos@ulg.ac.be \\

\pagebreak

\pagebreak {\large \bf  Figure Captions} \\

\par Fig.1 : Sketch of the behavior of the thermal diffusivity
for a
high-T$_c$ superconductor (continuous line) as a function of temperature. The
broken line ($\tilde{\alpha}^{+}$) is the extrapolated normal thermal diffusivity below
$T_c$; the shadowed area represents the superconducting phase contribution made of $\alpha_{e}^{s}$ and $\alpha_{ph}^{s}$ below $T_c$.  The line separating $\alpha_{e}^{s}$ and $\alpha_{ph}^{s}$ is merely indicative.

\par Fig.2 : (a) Normalized resistance of a DyBa$_2$Cu$_3$O$_7$ sample versus
temperature. (b) Normalized resistance of the sample when it is deoxygenized  to be
DyBa$_2$Cu$_3$O$_{6.3}$.  The data is shown when different magnetic fields are applied versus temperature \underline{in a
semi-log plot}.  The arrow indicates the antiferromagnetic transition.

\par Fig.3 : Comparison of the thermal conductivity of DyBa$_2$Cu$_3$O$_7$
($\bullet$) and DyBa$_2$Cu$_3$O$_{6.3}$ ($\circ$) versus temperature.

\par Fig.4 : Comparison of the thermal diffusivity of DyBa$_2$Cu$_3$O$_7$
($\bullet$) and DyBa$_2$Cu$_3$O$_{6.3}$ ($\circ$) versus the temperature
presented in a log-log plot.  Data below 50 K are represented by crosses.  The lines drawn correspond to Fig.1.  All data are normalized at $T=100$ K for better comparison with Fig.1.

\par Fig.5 : Plot of $\alpha_{oxy}-\alpha_{deox}$ ($\bullet$) obtained from Fig.4
versus the reduced temperature $\epsilon=|T-T_c|/T_c$ for comparing with the microwave loss
results obtained by Bonn et al. \cite{bonn} ($\circ$).  These measurements have been
rescaled by a multiplicative factor estimated at $100$ K for comparison.


\begin{thebibliography}{99}

\bibitem{ziman} D. Livanov and G. Fridman, Nuovo Cimento D {\bf 16},
325 (1994).

\bibitem{zeini} B. Zeini, A. Freimuth, B. B\"{u}chner, R. Gross, A.P.
Kampf, M. Kl\"{a}ser, and G. M\"{u}ller-Vogt, Phys. Rev. Lett. {\bf 82},
2175 (1999).

\bibitem{maneval} J.P. Maneval, F. Chibane, and R.W. Bland, Appl. Phys. Lett.
{\bf 61}, 339 (1992).


\bibitem{SUST} M. Ausloos and M. Houssa, Supercond. Sci. Technol.
{\bf 12}, R103 (1999).

\bibitem{zhang} Y. Zhang, N.P. Ong, P.W. Anderson, D.A. Bonn, R. Liang, and
W.N. Hardy, Phys. Rev. Lett. {\bf 86}, 890 (2001).

\bibitem{wolk}L. Terwordt and Th. W\"{o}lkhausen, Solid State Commun. {\bf 70},
83 (1989).

\bibitem{yu} R.C. Yu, M.B. Salomon, J.P. Lu, and W.C. Lee, Phys. Rev.
Lett. {\bf
69}, 1431 (1992); ibid. {\bf 71}, 1658 (1993).


\bibitem{houssa} M. Houssa, M. Ausloos, and S. Sergeenkov, J. Phys.: Condensed
Matter {\bf 8}, 2043 (1996); M. Houssa and M. Ausloos, Phys. Rev. B {\bf 51},
9372 (1995).

\bibitem{kirthley} J.R. Kirthley, C.C. Tsuei, K.A. Moler, J. Mannhart and H.
Hilgenkamp, {\it Symmetry and Pairing in Superconductors}, NATO Science Series
{\bf 63}, M. Ausloos and S. Krishinin eds. (Kluwer Academic Publishers,
Dordrecht, 1999) p.337.

\bibitem{scalapino} D.J. Scalapino, Phys. Rep.  {\bf 250}, 329 (1995).

\bibitem{bonn}D.A. Bonn, P. Dosanjh, R. Liang, and W.N. Hardy, Phys. Rev. Lett.
{\bf 68}, 2390 (1992).

\bibitem{krishana} K. Krishana, J.M. Harris, and N.P. Ong, Phys. Rev.
Lett. {\bf 75}, 3529 (1995).

\bibitem{bonn2} D.A. Bonn, S. Kamal, Kuan Zhank, Ruixing Liang, D.J. Baar, E.
Klein, and W.N. Hardy, Phys. Rev. B {\bf 50}, 4051 (1994).


\bibitem{MASS} S.A.Sergeenkov, V. V. Gridin,  and M. Ausloos, Z.
Phys. B 101, 565 (1996).

\bibitem{zhangand6} see \cite{zhang}, and references [1-6] therein

\bibitem{blatter} G. Blatter, M.V. Feigel'man, V.B. Geshkenbein, A.I.
Larkin, and
V.M. Vinokur, Rev. Mod. Phys. {\bf 66}, 1125 (1994).

\bibitem{cahill} D.G. Cahill and R.O. Pohl, Solid State Comm. {\bf 70}, 927
(1983).

\bibitem{epj} H. Bougrine, J.F. Geys, S. Dorbolo, R. Cloots, J.
Mucha, I. Nedkov,
and M. Ausloos, Eur. Phys. J. B {\bf 13}, 437 (2000).

\bibitem{kappaexp} H. Bougrine, Ph. D. Thesis (University of Li\`ege, 1995),
unpublished.

\bibitem{batonrouge} S. Dorbolo, H. Bougrine and M. Ausloos, Int. J.
Mod. Phys. B {\bf 12}, 3087 (1998).

\bibitem{cohn} J.L. Cohn, S.D. Peacor, and C. Uher, Phys. Rev. B {\bf 38},
2892 (1988).

\bibitem{allen} P.B. Allen, X.Du, and L. Mihaly, Phys. Rev. B {\bf 49},
9073 (1994).

\bibitem{dysynthese} R. Cloots, F. Auguste, A. Rulmont, N.
Vandewalle, and M. Ausloos, J. Mater. Res. {\bf 12}, 3199 (1997).

\bibitem{gmr} Qi Li and H.S. Wang, in {\it Nanocrystaline and Thin Film 
Magnetic  Oxidese}  I. Nedkov and M. Ausloos, Eds., NATO ASI Series 
vol. 72 (Kluwer, Dordrecht, 1999) p.133.

\bibitem{varma} B. Batlogg and C. Varma, Physics World {\bf 13}, 33 (2000).

\bibitem{ziman2} J.M. Ziman, {\it Electrons and Phonons, The theory of transport phenomena in solids}, (Clarendon Press, Oxford, 1967) p.320.

\bibitem{tyuterev} V.G. Tyuterev, P. Manca, and G. Mula, Physica C {\bf 297}, 32 (1998).

\bibitem{kato} H. Kato, K. Nara, M. Okaji, M. Hirabayashi, and H. Ihara, Czech.
J. Physics {\bf 46}, 1179 (1996).

\bibitem{calzona} V. Calzona, M.R. Cimberle, C. Ferdeghini, M. Putti,
C. Rizzuto,
and A.S. Siri, Europhys. Lett. {\bf 13}, 181 (1990).


\bibitem{castel} S. Castellazzi, M.R. Cimberle, C. Ferdeghini, E. Giannini,
G. Grasso, D. Marrè, M. Putti, and A.S. Siri, Physica C {\bf 273}, 314
(1997).

\bibitem{wu} X.D. Wu, J.G. Fanton, G.S. Kino, S. Ryu, D.B. Mitzi, and A.
Kapitulnik, Physica C {\bf 218}, 417 (1993).


\bibitem{ikebe} M. Ikebe, H. Fujishiro, T. Naito, and K. Noto, J.
Phys. Soc. Jpn {\bf 63}, 3107 (1994).


\end{thebibliography}
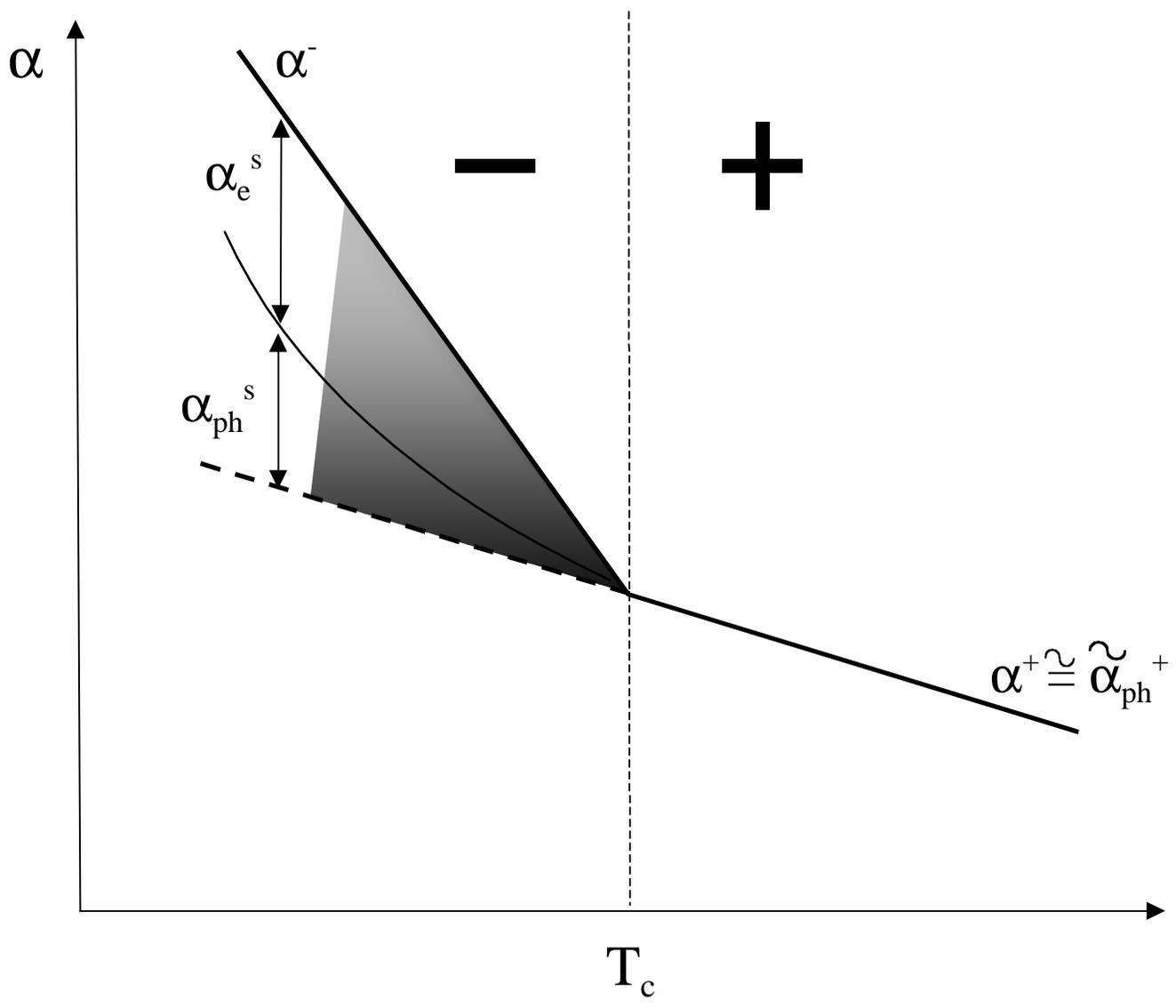
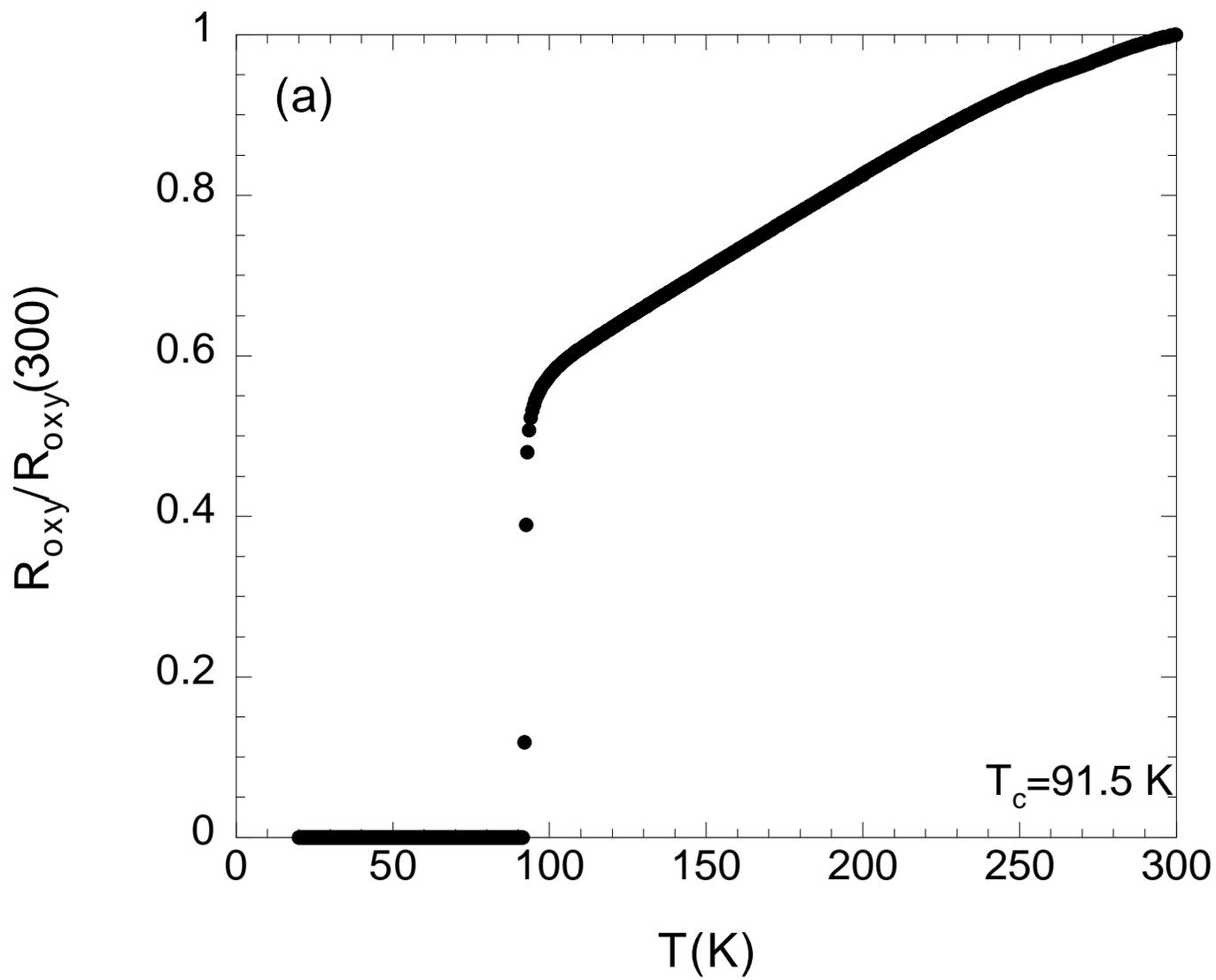
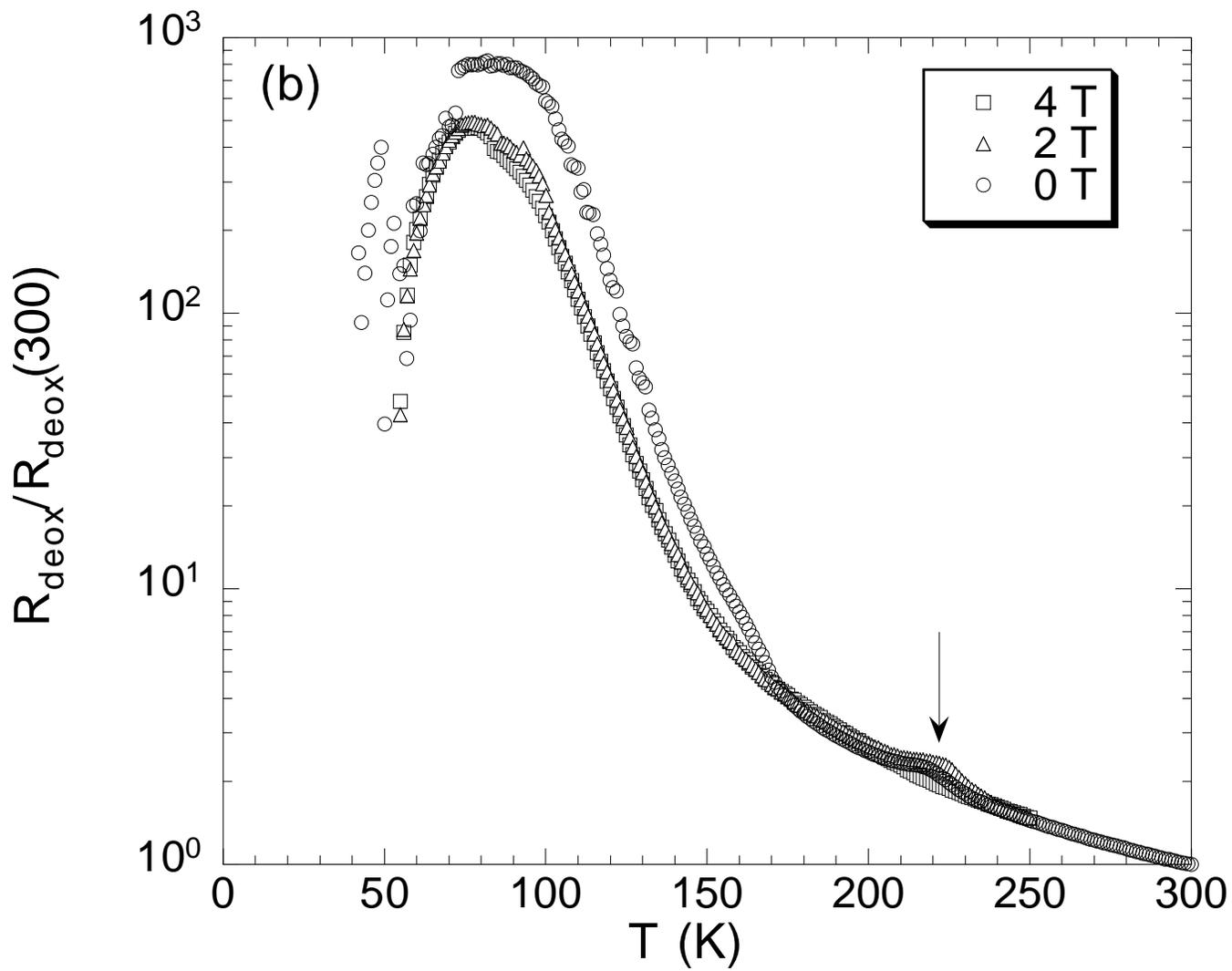
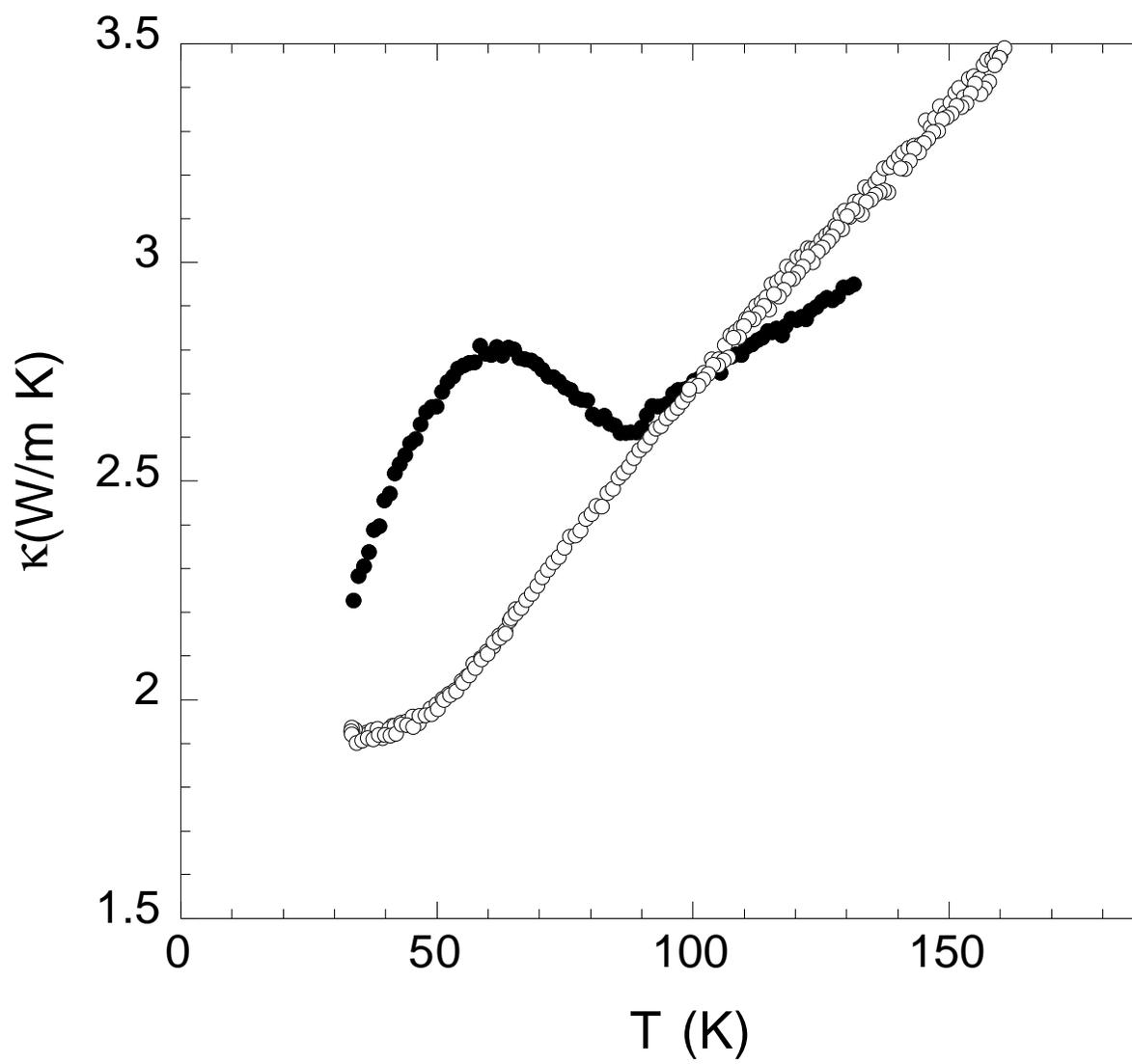
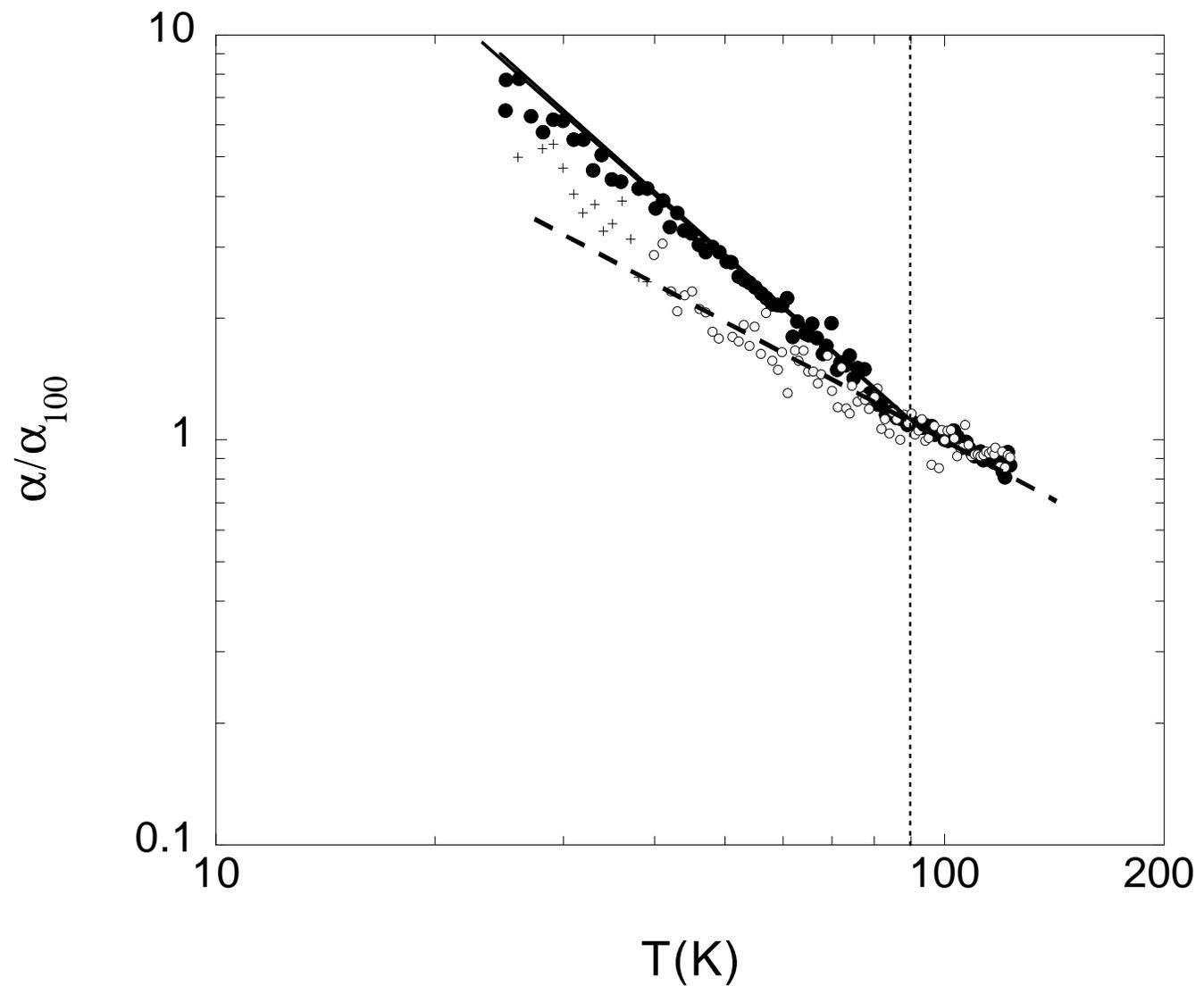
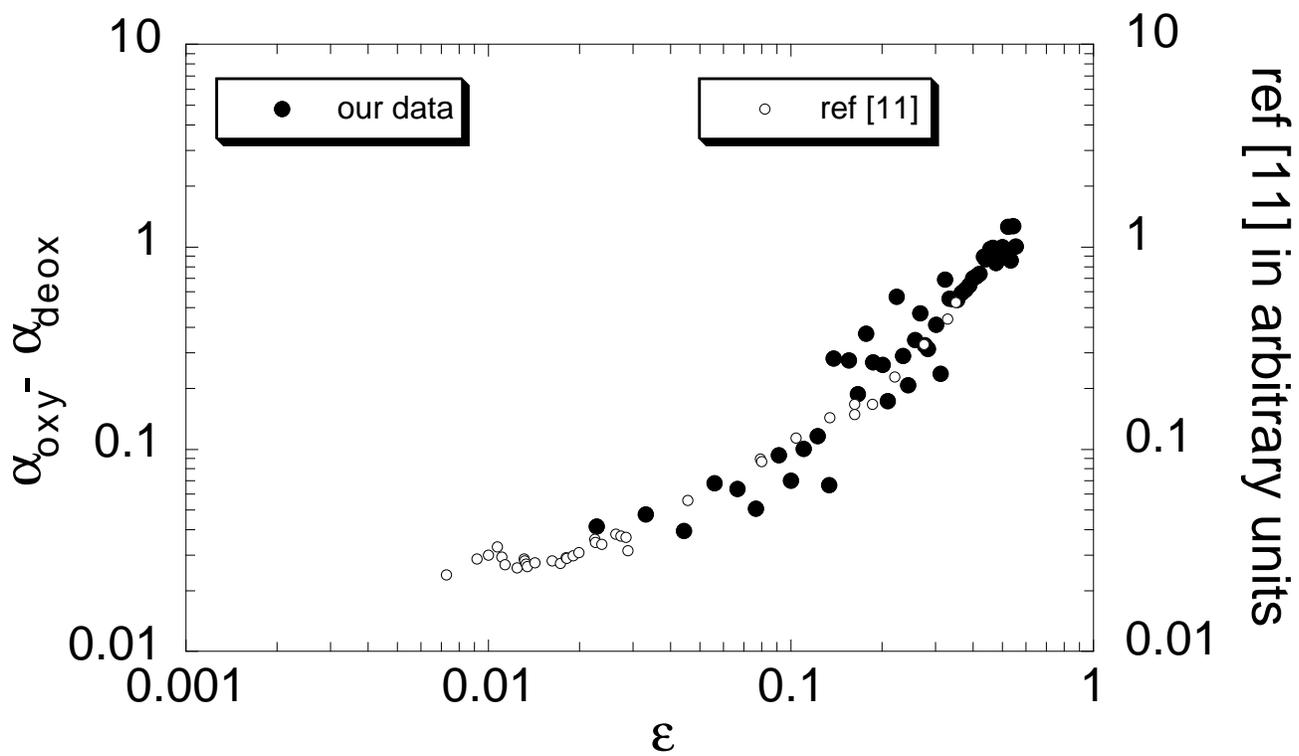
\end{document}